\documentclass{article}
\usepackage{dcase2020,amsmath,graphicx,url,times,booktabs, tabularx}

\def\tthline{\noalign{\hrule height 1.4pt}}

\newcolumntype{?}{!{\vrule width 1pt}}
\usepackage{hyperref}
\hypersetup{
    colorlinks=true,
    linkcolor=blue,
    filecolor=magenta,      
    urlcolor=cyan,
}

\title{Model selection for deep audio source \\ separation via clustering analysis}

\name{Alisa Liu, Prem Seetharaman, Bryan Pardo}
\address{Northwestern University\\
      Computer Science Department\\
      Evanston, IL}

\begin{document}

\maketitle

\begin{abstract}
Audio source separation is the process of separating a mixture into isolated sounds from individual sources. Deep learning models are the state-of-the-art in source separation, given that the mixture to be separated is similar to the mixtures the deep model was trained on. This requires the end user to know enough about each model's training to select the correct model for a given audio mixture. In this work, we propose a confidence measure that can be broadly applied to any clustering-based separation model. The proposed confidence measure does not require ground truth to estimate the quality of a separated source. We use our confidence measure to automate selection of the appropriate deep clustering model for an audio mixture. Results show that our confidence measure can reliably select the highest-performing model for an audio mixture without knowledge of the domain the audio mixture came from, enabling automatic selection of deep models.
\end{abstract}

\section{Introduction}
\label{sec:intro}
Audio source separation is the process of separating a mixture into isolated sounds from individual sources. It enables tasks where it would be valuable to attend to or manipulate individual sounds in a mixture. Examples include automatic speech recognition with multiple speakers, music manipulation, and content-based audio search. 

Deep learning models are state-of-the-art for source separation, given that the mixture to be separated is similar to the mixtures the model was trained on \cite{ward2018sisec}. 
Unfortunately, a deep separation model trained on one domain does not generalize well to others. For example, using a model trained on music examples to isolate a single voice from a recording of multiple concurrent voices will not produce usable results. Thus, the end user must know enough about each model's training to select the appropriate model for a given audio mixture. This limits how models can be deployed and imposes a bottleneck on adding source separation to an automation chain.

\begin{figure}
    \centering
    \includegraphics[width=1.0\linewidth]{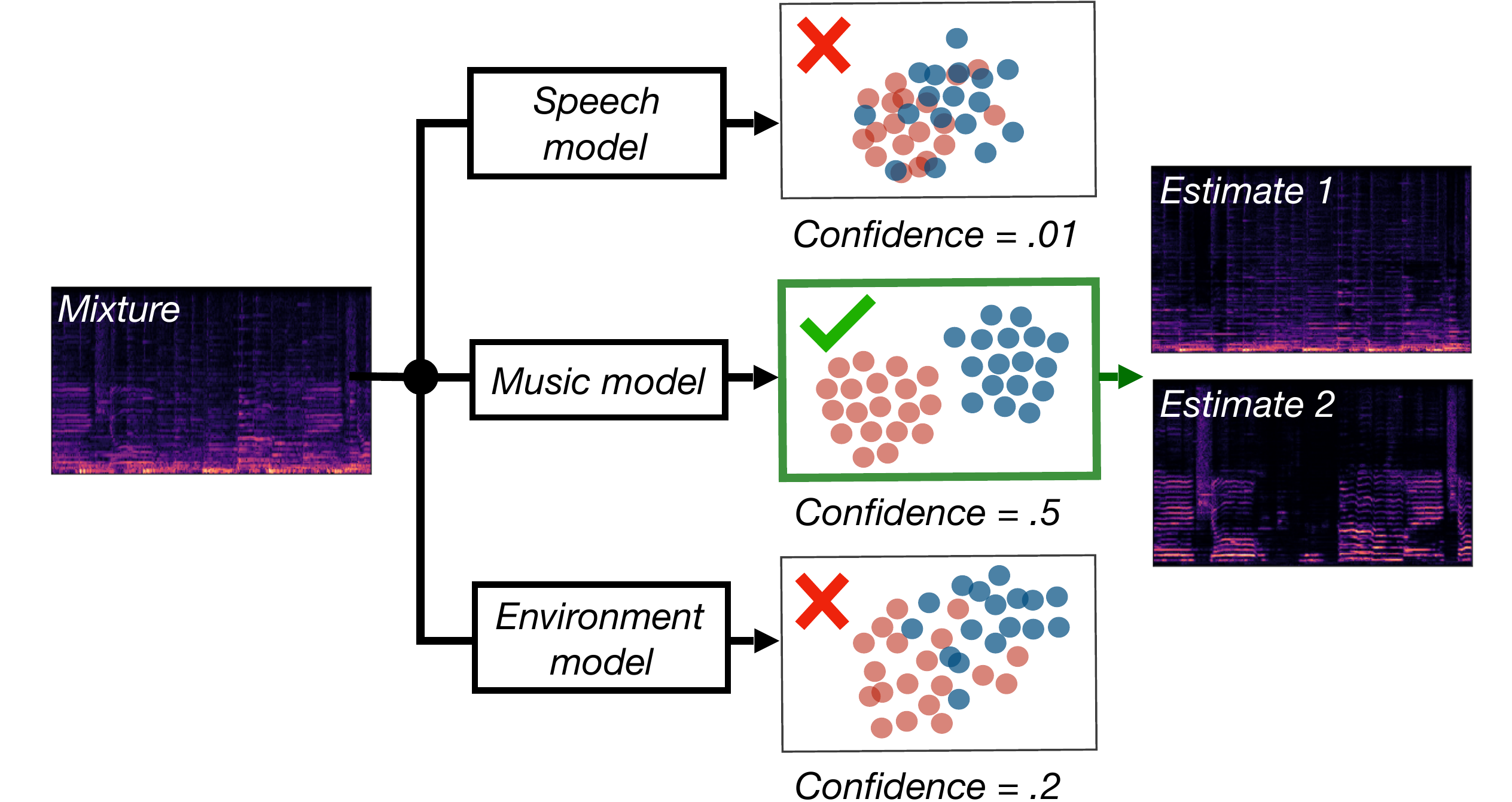}
    \caption{Overview of our system. The mixture is passed through three deep source separation models, each trained on a different domain (speech, music, environmental sounds). We use the confidence measure to select which of the three model outputs to use as the final estimates.}
    \label{fig:schematic}
\end{figure}

A method to automatically select the best model for the current mixture would transform the range of applications where source separation can be applied. For example, imagine a hearing aid that automatically switches models when the user moves from an outdoor construction site (a model trained to separate speech from environmental noise) to an indoor restaurant (a model trained to separate speech from speech). Moreover, simply knowing when no source separation model available for the current audio scene is useful. When source separation would produce inappropriate output that might be worse than not applying separation at all, a device could simply turn off source separation. 

In this work, we develop a confidence measure that can be applied to systems that perform clustering on embedding spaces to do source separation, as embodied by the family of \textit{deep clustering} \cite{Hershey2016deepclustering} methods. These have been among the most successful source separation approaches in recent years \cite{wang2018alternative, luo2016deep}. We use this confidence measure to automatically select the model output with the best predicted separation quality.  A system overview is shown in Figure \ref{fig:schematic}. 





\begin{figure*}[ht]
\centering
\includegraphics[scale=0.35]{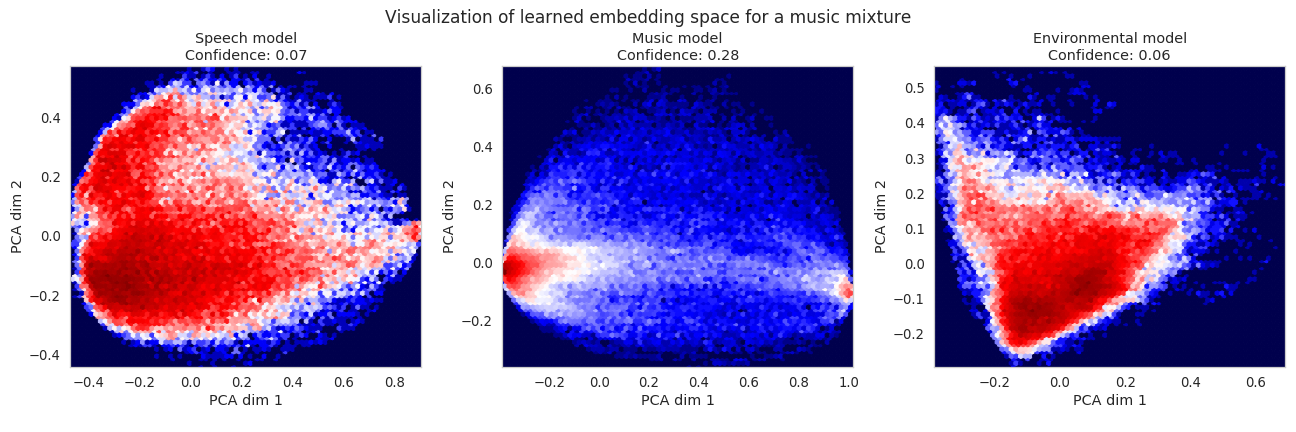}
\caption{2D visualization of the embedding of a single music mixture produced by three different models. Visually, the embedding of the music model is far more clusterable than the speech and environmental model. Therefore the mixture may be easier to separate with the music model.}
\label{fig:visualization}
\end{figure*}

\section{Prior Work}
\label{sec:prior_work}
Kim et al.~\cite{kim2017collaborative} produces a selection method for speech enhancement by training an ensemble of autoencoders and selecting the model with the lowest reconstruction error. While speech enhancement is related to source separation, the underlying assumption is that the desired signal constitutes the majority of the audio. Since, source separation requires removal of large portions of the input audio, low reconstruction error does not correspond to good source separation.

In computer vision, failure detection has employed a measure of confidence that is learned during training \cite{devries2018learning} or by comparing the the input image to the training data \cite{subramanya2017confidence, hendrycks2016baseline}. Our work is for audio, does not require a specific model for error estimation, and does not need access to the training distribution. The clusterability of audio representations was considered in \cite{pishdadian2018MCFT}, where clusterability was computed when the information about true label assignments is already known. Here we use unsupervised clusterability measures where the label assignments are estimated.


There has been work in combining multiple algorithms for source separation \cite{mcvicar2016learning, driedger2015extracting, liutkus2014kam}. In particular, \cite{manilow2017predicting} trained a separate deep net to estimate source separation quality in order to guide switching between separation algorithms. In contrast, our confidence measure does not require any training. Moreover, all of these ensemble methods were applied to multiple methods that work for one application domain (e.g. two separation approaches, both applied to speech separation) whereas our method chooses which model is the most appropriate for a given mixture across multiple domains. Our confidence measure builds on the one in \cite{seetharaman2019bootstrapping}, which was designed to predict the performance of a direction-of-arrival algorithm for source separation. In that work, the confidence measure was designed particularly for speech mixtures and required the clustering to be based on Gaussian Mixture Models. The confidence measure here can be applied anywhere that embedding spaces are clustered. In this paper we apply it to an ensemble of deep clustering networks.

\section{Proposed Method}
\label{sec:method}

\subsection{Deep clustering}

In deep clustering, a neural network is trained to map each time-frequency bin in a magnitude spectrogram of an audio mixture to a higher-dimensional embedding, such that bins that primarily contain energy from the same source are near each other and bins whose energy primarily come from different sound sources are far from each other. Given a good mapping, the assignment of bin to source can then be determined by a simple clustering method, such as K-Means. All members of the same cluster are assigned to the same source. Because deep clustering performs separation via clustering, our method relies on an analysis of the embedding space produced by deep clustering to establish a confidence measure. 

\subsection{Confidence measure}
The core insight behind the confidence measure is that the distribution of the embedded time-frequency points of a mixture is predictive of the performance of the algorithm.  Figure \ref{fig:visualization} shows a visualization of the confidence measure as applied to the distribution of points in a mixture produced by three trained deep clustering networks, each trained on a different domain. The input is a music mixture. The speech (left) and environmental (right) models return distributions with no clear clusters. The music model (middle) shows a more clusterable distribution, which is reflected by a higher confidence score.
The confidence measure $C(X)$ combines the silhouette score $S(X)$ and posterior strength $P(X)$ through multiplication so that it is high only when both are high. That is, $C(X)=S(X)P(X)$.

\subsubsection{Silhouette score}

The silhouette score \cite{rousseeuw1987silhouettes} captures how far apart clusters are (intercluster distance) and how dense they are (intracluster distance). Define $X$ as the set of embeddings for every time-frequency point in an audio mixture, where $x_i$ is the embedding of one point. $X$ is partitioned into $K$ clusters $C_k$: $X = \bigcup_{k=1}^K C_k$. Given a data point $x_i$ assigned to cluster $C_k$:
\begin{align}
a(x_i) &= \frac{1}{|C_k| - 1} \sum_{\substack{x_j \in C_k,\\ x_i \neq x_j}} d(x_i, x_j) \\
b(x_i) &= \min_{\ell \neq k} \frac{1}{|C_\ell|} \sum_{x_j \in C_\ell} d(x_i, x_j)
\end{align}
Here, the intracluster distance $a(x_i)$ is the mean distance (using a distance function $d$) between $x_i$ and all other points in $C_k$, and the intercluster distance $b(x_i)$ is the mean distance between $x_i$ and all the points in the nearest cluster $C_\ell$. Compute the \textit{silhouette score} of $x_i$
\begin{equation}
    \label{eq:chap3:silhouette}
	s\left(x_i\right) = \frac{b\left(x_i\right) - a\left(x_i\right)}{\max \left\{ a(x_i), b\left(x_i\right)\right\}}
\end{equation}
Note $s(x_i)$ ranges from $-1$ to $1$. In the silhouette score, every point in every cluster must be compared to every point in every other cluster. Since a time-frequency representation can easily contain upwards of a million points, computing the silhouette score for every point in embedding space is intractable. Therefore, we select a constant number of points ($N=1000$) from the loudest $1\%$ of all time-frequency bins. This is because the assignment of louder time-frequency bins is more important perceptually than softer time-frequency bins. We take the mean silhouette score across the sample to estimate the silhouette score $S(X)$ for the mixture.

\subsubsection{Posterior strength}

For every point $x_i$ in a dataset $X$, the clustering algorithm, soft K-Means, produces $\gamma_{ik} \in [0, 1]$, which indicates the membership of the point $x_i$ in some cluster $C_k$, also called the \textit{posterior} of the point $x_i$ in regards to the cluster $C_k$. The closer that $\gamma_{ik}$ is to $0$ (not in the cluster) or $1$ (in the cluster), the more sure the assignment of that point. We compute the \textit{posterior strength} of $x_i$ as follows:
\begin{equation}
    \label{eq:chap3:posterior_confidence}
    P(x_i) = \frac{K \left(\max\limits_{k \in [0, ..., K]} \gamma_{ik}\right) - 1}{K - 1}
\end{equation}
The equation maps points that have a max posterior of $\frac{1}{K}$ (equal assignment to all clusters) to $0$, and points that have a max posterior of $1$ to $1$. The overall posterior strength $P(X)$ is the mean $P(x_i)$ for the loudest $1\%$ of time-frequency bins. 

We apply this in conjunction with the silhouette score to predict separation performance across multiple domains. An implementation is available\footnote{ \href{https://git.io/Jf3y5}{https://git.io/Jf3y5}} and integrated into \textit{nussl} \cite{manilow2018nussl}.

\section{Experimental design}
\label{sec:experiments}

Our experiments are designed to investigate two facets of the proposed confidence measure. The first is whether or not the confidence measure correlates with
ground truth separation performance, measured via signal-to-distortion ratio (SDR) \cite{roux2018sdr}. The second is whether we can use the confidence measure to choose the appropriate deep clustering model given an audio mixture that comes from any of the three domains we consider: environmental, speech, and musical sounds.

Within each audio domain, we consider a specific task. For speech, the task was to separate two speakers talking simultaneously. For music, the task was to separate singing voice from accompaniment (drums, bass, and other instruments). For environmental sounds, the task was to separate two environmental sounds from one another (e.g. isolating a dog bark from a car horn). For each domain, we trained a source separation network with a deep clustering objective. Each network is identical, consisting of 2 BLSTM layers with 300 hidden units in both directions. The embedding size of each network is 20 with sigmoid activation and is trained for 80 epochs with the Adam optimizer \cite{kingma2014adam} (learning rate of 2e-4).

\begin{figure}[t]
  \centering
  \centerline{\includegraphics[scale=0.40]{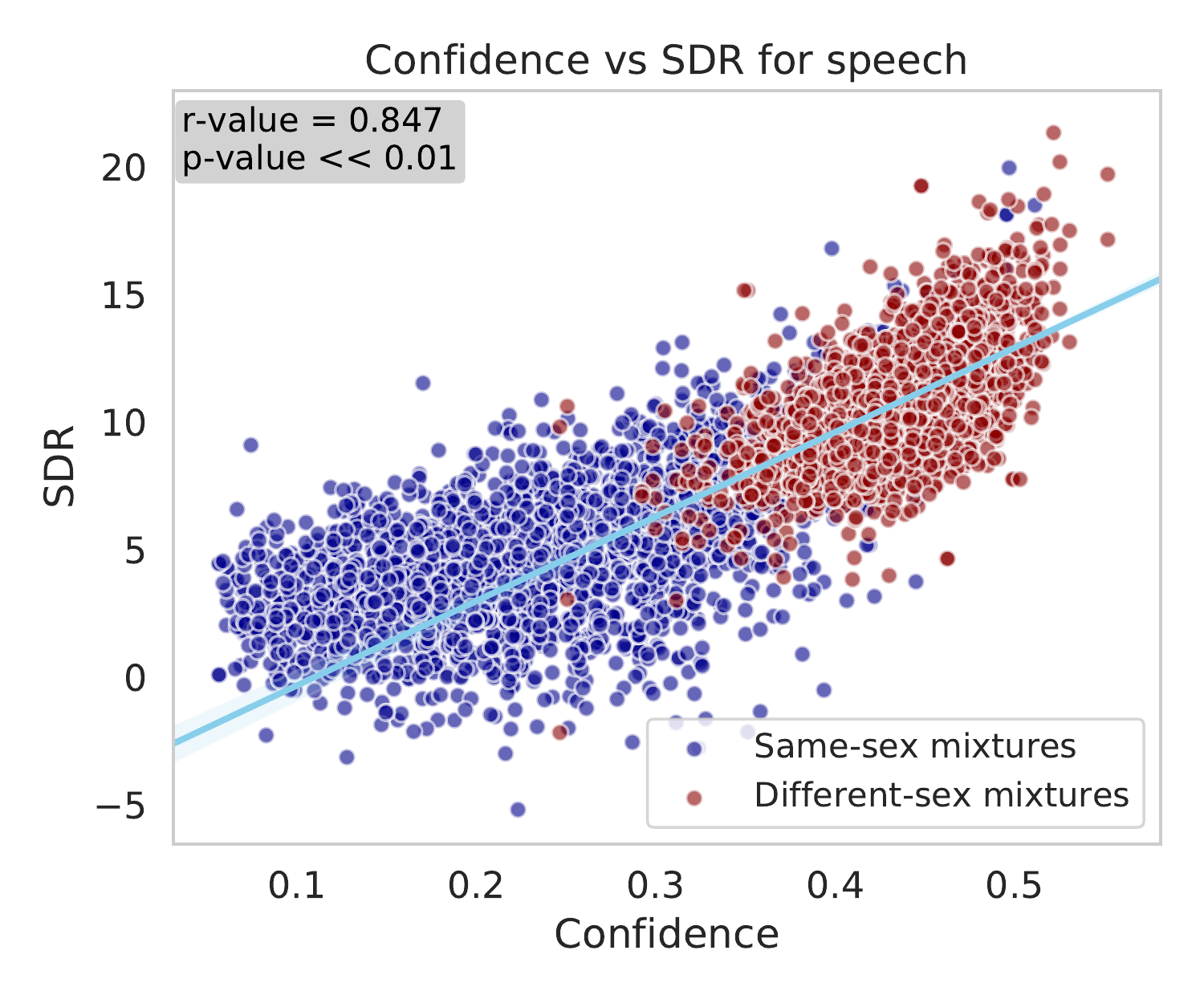}}
\caption{Relationship between confidence measure and actual performance based on SDR. Each dot represents a separated source by the speech model. The blue line is the line of best fit found via linear regression.}
\label{fig:correlation}
\end{figure}

We create a dataset for each domain. The speech dataset is constructed using the Wall Street Journal dataset \cite{garofolo1993WSJ0}, the music dataset from the MUSDB dataset \cite{musdb18}, and the environmental dataset from the UrbanSound8k dataset \cite{Salamon:UrbanSound:ACMMM:14}. 

The Wall Street Journal dataset consists of utterances from 119 speakers. These speakers are split into train, validation, and test sets by prior work \cite{Hershey2016deepclustering}. The test set contains utterances from speakers that are not in either the train or validation set (referred to as the open speaker set). We use these splits when creating our speech mixtures dataset.

The MUSDB music dataset has four isolated stems for each song: vocals, drums, bass, and other. We define the accompaniment as the sum of the drums, bass, and other stems. The network is trained to separate the vocals from the accompaniment given the music mixture. MUSDB comes split into train, validation, and test sets.

The UrbanSound8k dataset has 10 sound classes: air conditioner, car horn, children playing, dog bark, drilling, engine idling, gun shot, jackhammer, siren and street music. We use 5 classes which do not contain stationary noise: car horn, dog bark, gun shot, jackhammer, siren. The other sound classes are polyphonic or multisource (e.g. children playing contains cars in the background) and are excluded from experiments. UrbanSound8k comes split into 10 folds. We use folds 1-8 for training sources, 9 for validation, and 10 for testing.

We downsample all audio to $16$ kHz sample rate to reduce computational costs. For each dataset, we made 5-second
mixtures with 2 sources each using Scaper \cite{salamon2017scaper}. For Wall Street Journal and UrbanSound8k, the $2$ sources were mixed at a random signal-to-noise ratio (SNR) between $-2.5$ and $2.5$ dB, relative to each other. For MUSDB, the vocals were mixed at an SNR of $10$ dB relative to the accompaniment. For each domain, we made $20000$ mixtures for training, $5000$ for validation, and $3000$ for testing.

\begin{table}[]
\centering
\footnotesize
\begin{tabular}{ccccc}
 & & & \textbf{True domain} &  \\
\rotatebox[origin=c]{90}{\textbf{Predicted domain}} & 
 \multicolumn{3}{c}{
    \begin{tabular}{l ?? c|c|c}
    \textbf{Domain} & \textbf{Speech}   & \textbf{Music}    & \textbf{Environ.}  \\
    \hline \hline
    \textbf{Speech} & 2186 & 25 & 31 \\ \hline
    \textbf{Music} & 717 & 2950 & 1235 \\ \hline
    \textbf{Environ.} & 97 & 25 & 1734 \\ \hline \hline
    \textbf{Total} & 3000 & 3000 & 3000
    \end{tabular}
} \\

\label{tab:confusion}

\end{tabular}
\caption{Confusion matrix comparing the true domain with the domain-specific model selected by the confidence measure. We see that the ensemble fairly reliably picks the correct model for each domain (the diagonal of the confusion matrix).}
\end{table}

\begin{table}[]
    \centering
    \footnotesize
    \begin{tabular}{l ?? c|c|c}
    \textbf{Approach} & \textbf{Speech}   & \textbf{Music}    & \textbf{Environ.}  \\ \tthline
    Oracle ensemble & 8.3 & 6.5 & 12.2 \\ \hline
    Confidence ensemble & 7.6 & 6.4 & 10.5 \\ \hline
    Random ensemble & 4.8 & 4.2 & 2.8 \\ \hline \hline
    Speech model & 8.2 & 2.0 & 3.0 \\ \hline
    Music model & 1.4 & 6.5 & 2.5 \\ \hline
    Environ. model & 2.1 & 1.7 & 11.9 \\
    \end{tabular}
    \caption{Performance (mean SDR) of different approaches on test datasets from three domains. Higher values are better. The top three rows correspond to the different ensembles. The bottom three rows show the result of using a single model across all the 
    domains.}
    \label{tab:ensemble}
\end{table}

\section{Results}

First, we investigate whether the confidence measure correlates well to ground-truth performance in terms of SDR for a single model. Figure \ref{fig:correlation} shows a clear relationship between confidence and SDR for speech mixtures separated by the speech model. We see that both confidence and performance are a function of the mixture \textit{type}: same-sex mixtures are harder to separate. This was also observed by \cite{ditter2019influence}, where they used 
domain-specific knowledge about the psycho-acoustic differences between male and female speech. Here, we have uncovered that same relationship only by analyzing the clusterability of embeddings. We observe similarly strong correlations on the music (r-value of $.46$) and environmental (r-value of $.70$) domains. 

Next, we compare the confidence-based ensemble approach with other approaches to separating each dataset. The top three rows of Table \ref{tab:ensemble} show the performance of three ensemble approaches. \textit{Oracle}, which switches with knowledge of the best system, is the upper bound. \textit{Random}, which selects randomly with equal probability, is the lower bound. \textit{Confidence} uses our confidence measure to select the model: in this method, all three models are run and confidence measures are computed for each separation, and the output with the highest confidence is chosen. 


We observe that the confidence-based ensemble significantly outperforms the random ensemble. In the case of music mixtures, the confidence-based model achieves almost oracle performance. In Table \ref{tab:confusion}, we show a confusion matrix to compare the true domain with the predicted domain. The precision of picking the best model when the best model is speech, music, and environmental sound are $.97$, $.60$, and $.93$, respectively. The recall rates are $.72$, $.98$, and $.57$. 

\section{Conclusion}
We have presented a method for constructing ensembles of deep clustering models by using off-the-shelf clustering analysis techniques such as the silhouette score and posterior analysis. This confidence measure can be applied to ensembles of any clustering-based separation algorithms. Our work can be integrated into interfaces that use source separation by easing the burden on the user for selecting an appropriate model, as well as alerting them to system failure.

\small
\bibliographystyle{IEEEbib}
\bibliography{strings,refs}

\begin{thebibliography}{10}

\bibitem{ward2018sisec}
Dominic Ward, Russel~D Mason, Chungeun Kim, Fabian-Robert St{\"o}ter, Antoine
  Liutkus, and Mark Plumbley,
\newblock ``Sisec 2018: State of the art in musical audio source
  separation-subjective selection of the best algorithm,''
\newblock 2018.

\bibitem{Hershey2016deepclustering}
{John R.} Hershey, Zhuo Chen, Jonathan {Le Roux}, and Shinji Watanabe,
\newblock ``Deep clustering: Discriminative embeddings for segmentation and
  separation,''
\newblock in {\em IEEE International Conference on Acoustics, Speech and Signal
  Processing (ICASSP)}, 2016, pp. 31--35.

\bibitem{wang2018alternative}
Zhong-Qiu Wang, Jonathan Le~Roux, and John~R Hershey,
\newblock ``Alternative objective functions for deep clustering,''
\newblock in {\em IEEE International Conference on Acoustics, Speech and Signal
  Processing (ICASSP)}, 2018.

\bibitem{luo2016deep}
Yi~Luo, Zhuo Chen, John~R Hershey, Jonathan Le~Roux, and Nima Mesgarani,
\newblock ``Deep clustering and conventional networks for music separation:
  Stronger together,''
\newblock in {\em IEEE International Conference on Acoustics, Speech and Signal
  Processing (ICASSP)}, 2017, pp. 61--65.

\bibitem{kim2017collaborative}
Minje Kim,
\newblock ``Collaborative deep learning for speech enhancement: A run-time
  model selection method using autoencoders,''
\newblock in {\em IEEE International Conference on Acoustics, Speech and Signal
  Processing (ICASSP)}, 2017, pp. 76--80.

\bibitem{devries2018learning}
Terrance DeVries and Graham~W Taylor,
\newblock ``Learning confidence for out-of-distribution detection in neural
  networks,''
\newblock in {\em IEEE International Conference on Multimedia and Expo (ICME)},
  2018.

\bibitem{subramanya2017confidence}
Akshayvarun Subramanya, Suraj Srinivas, and R~Venkatesh Babu,
\newblock ``Confidence estimation in deep neural networks via density
  modelling,''
\newblock in {\em IEEE International Conference on Multimedia and Expo (ICME)},
  2017.

\bibitem{hendrycks2016baseline}
Dan Hendrycks and Kevin Gimpel,
\newblock ``A baseline for detecting misclassified and out-of-distribution
  examples in neural networks,''
\newblock in {\em International Conference on Learning Representations (ICML)},
  2017.

\bibitem{pishdadian2018MCFT}
Fatemeh Pishdadian and Bryan Pardo,
\newblock ``Multi-resolution common fate transform,''
\newblock {\em Transactions on Audio, Speech, and Language Processing}, 2018.

\bibitem{mcvicar2016learning}
Matt {McVicar}, Raul {Santos-Rodríguez}, and Tijl {De Bie},
\newblock ``Learning to separate vocals from polyphonic mixtures via ensemble
  methods and structured output prediction,''
\newblock in {\em IEEE International Conference on Acoustics, Speech and Signal
  Processing (ICASSP)}, 2016, pp. 450--454.

\bibitem{driedger2015extracting}
Jonathan Driedger and Meinard M{\"u}ller,
\newblock ``Extracting singing voice from music recordings by cascading audio
  decomposition techniques,''
\newblock in {\em IEEE International Conference on Acoustics, Speech and Signal
  Processing (ICASSP)}, 2015, pp. 126--130.

\bibitem{liutkus2014kam}
Antoine Liutkus, Derry Fitzgerald, Zafar Rafii, Bryan Pardo, and Laurent
  Daudet,
\newblock ``Kernel additive models for source separation,''
\newblock {\em IEEE Transactions on Signal Processing}, vol. 62, no. 16, pp.
  4298--4310, 2014.

\bibitem{manilow2017predicting}
Ethan Manilow, Prem Seetharaman, Fatemeh Pishdadian, and Bryan Pardo,
\newblock ``Predicting algorithm efficacy for adaptive multi-cue source
  separation,''
\newblock in {\em IEEE Workshop on Applications of Signal Processing to Audio
  and Acoustics (WASPAA)}, 2017.

\bibitem{seetharaman2019bootstrapping}
Prem Seetharaman, Gordon Wichern, Jonathan Le~Roux, and Bryan Pardo,
\newblock ``Bootstrapping single-channel source separation via unsupervised
  spatial clustering on stereo mixtures,''
\newblock in {\em ICASSP 2019-2019 IEEE International Conference on Acoustics,
  Speech and Signal Processing (ICASSP)}, 2019, pp. 356--360.

\bibitem{rousseeuw1987silhouettes}
Peter~J Rousseeuw,
\newblock ``Silhouettes: a graphical aid to the interpretation and validation
  of cluster analysis,''
\newblock {\em Journal of computational and applied mathematics}, vol. 20, pp.
  53--65, 1987.

\bibitem{manilow2018nussl}
Ethan Manilow, Prem Seetharaman, and Bryan Pardo,
\newblock ``The northwestern university source separation library,''
\newblock in {\em Proceedings of the 19th International Society of Music
  Information Retrieval Conference (ISMIR 2018)}, 2018.

\bibitem{roux2018sdr}
Jonathan Le~Roux, Scott Wisdom, Hakan Erdogan, and John~R Hershey,
\newblock ``{SDR} -- half-baked or well done?,''
\newblock in {\em IEEE International Conference on Acoustics, Speech and Signal
  Processing (ICASSP)}, 2019.

\bibitem{kingma2014adam}
Diederik~P Kingma and Jimmy Ba,
\newblock ``Adam: A method for stochastic optimization,''
\newblock {\em arXiv preprint arXiv:1412.6980}, 2014.

\bibitem{garofolo1993WSJ0}
John Garofolo, D~Graff, D~Paul, and D~Pallett,
\newblock ``Csr-i ({WSJ0}) complete {LDC93S6A},''
\newblock {\em Web Download. Philadelphia: Linguistic Data Consortium}, 1993.

\bibitem{musdb18}
Zafar Rafii, Antoine Liutkus, Fabian-Robert St{\"o}ter, Stylianos~Ioannis
  Mimilakis, and Rachel Bittner,
\newblock ``The {MUSDB18} corpus for music separation,'' 2017.

\bibitem{Salamon:UrbanSound:ACMMM:14}
Justin Salamon, Christopher Jacoby, and Juan~Pablo Bello,
\newblock ``A dataset and taxonomy for urban sound research,''
\newblock in {\em ACM international conference on Multimedia}, 2014, pp.
  1041--1044.

\bibitem{salamon2017scaper}
Justin Salamon, Duncan MacConnell, Mark Cartwright, Peter Li, and Juan~Pablo
  Bello,
\newblock ``Scaper: A library for soundscape synthesis and augmentation,''
\newblock in {\em IEEE Workshop on Applications of Signal Processing to Audio
  and Acoustics (WASPAA)}, 2017.

\bibitem{ditter2019influence}
David Ditter and Timo Gerkmann,
\newblock ``Influence of speaker-specific parameters on speech separation
  systems,''
\newblock {\em ISCA Interspeech, Graz, Austria}, pp. 4584--4588, 2019.

\end{thebibliography}

\end{document}